\documentclass[twocolumn,showpacs,preprintnumbers,amsmath,amssymb]{revtex4}

\usepackage{graphicx}
\usepackage{dcolumn}
\usepackage{bm}

\begin{document}

\title{Spin configurations in circular and rectangular vertical quantum dots in a magnetic field: Three-dimensional self-consistent simulation}

\author{Dmitriy V. Melnikov, Philippe Matagne, and Jean-Pierre Leburton}
\affiliation{
Beckman Institute for Advanced Science and Technology\\
University of Illinois at Urbana-Champaign\\
401 North Mathews Avenue\\
Urbana, IL 61801
}

\author{D.G. Austing and G. Yu}
\affiliation{
Institute for Microstructural Sciences\\
National Research Council of Canada\\
Montreal Road\\
Ottawa, Ontario K1A 0R6, Canada
}

\author{S. Tarucha}
\altaffiliation[Also at the ]{Department of Applied Physics,
University of Tokyo,
Hongo, Bunkyo-ku, Tokyo 113-0033, Japan
}
\affiliation{
Tarucha Mesoscopic Correlation Project\\
ERATO, JST\\
Atsugi-shi, Kanagawa 243-0198, Japan}

\author{John Fettig and Nahil Sobh}
\affiliation{
National Center for Supercomputing Applications\\
University of Illinois at Urbana-Champaign\\
401 North Mathews Avenue\\
Urbana, IL 61801
}

\date{\today}


\begin{abstract}
The magnetic field dependence of the electronic properties of \textit{real} single vertical quantum dots in circular and rectangular mesas is investigated within a full three-dimensional multiscale self-consistent approach without any {\it \'a priori} assumptions about the shape and strength of the confinement potential. The calculated zero field electron addition energies are in good agreement with available experimental data for both mesa geometries. Charging diagrams in a magnetic field for number of electrons up to five are also computed. Consistent with the experimental data, we found that the charging curves for the rectangular mesa dot in a magnetic field are flatter and exhibit less features than for a circular mesa dot. Evolution of the singlet-triplet energy separation in the two electron system for both dot geometries in magnetic field was also investigated. In the limit of large field, beyond the singlet-triplet transition, the singlet-triplet energy difference continues to become more negative in a circular mesa dot without any saturation within the range of considered magnetic fields whilst it is predicted to asymptotically approach zero for the rectangular mesa dot. This different behavior is attributed to the symmetry "breaking" that occurs in the singlet wave-functions in the rectangular mesa dot but not in the circular one.
\end{abstract}

\pacs{73.21.-b}

\maketitle

\section{Introduction}
\label{Intro}

In recent years the electronic properties of semiconductor quantum dots (QDs) have attracted considerable attention because these zero-dimensional objects exhibit properties characteristic of atomic few-electron systems. QDs are also regarded as promising candidates for future device applications \cite{1}, not least of which is the possibility of realizing a scalable quantum computer \cite{2}. A considerable amount of theoretical and experimental research has been devoted to understanding the electronic properties of cylindrical mesa structures within which the QD has a shape of a thin disc. These disc-shaped QDs are usually formed from double barrier structures made in to small mesas and surrounded by a gate electrode. The number of electrons is fully tunable starting from zero~\cite{3}. In such systems three-dimensional quantization coupled with rotational symmetry leads to the existance of a "shell structure" in the electron addition energy, {\it i.e.} for certain electron numbers the electron addition energies show pronounced maxima. Spin effects also lead to the appearance of smaller secondary peaks in the mid-shell regions, in accordance with Hund's first rule \cite{3a}. 

In contrast to real atoms, electrons in vertical QDs occupy a sizable region of space of the order of 100 nm. This allows utilization of a relatively small and readily accessible magnetic field to probe their electronic structure and also to induce transitions between the various electronic states as the number of particles and the strength of field are varied. However, perfect circular symmetry of the QD is difficult to achieve and maintain in real scalable electronic circuits. Ideal atomistic features are also very sensitive to any kind of disruption in the system \cite{3b}. This dictates the necessity of understanding the response of the electronic structure to various geometric deformations. 

Compared with circular QDs which have been studied by a variety of methods \cite{3a}, QDs in rectangular mesas (hence forth called rectangular QDs for brevity) have attracted considerably less attention both experimentally and theoretically. Invariably, most calculations assume a pure two-dimensional confinement potential. Frequently \cite{4,6,6a,8}, a {\it model} two-dimensional anisotropic harmonic oscillator potential $(1/2)m^*(\omega_x^2x^2+\omega_y^2y^2)$ characteristic of an elliptic QD with confinement strengths $\omega_x$ and $\omega_y$ in the $x$ and $y$ directions respectively, is used in theoretical calculations for this purpose. Simulations conducted with such a potential \cite{4} showed that it is not straightforward to match the experimental geometric aspect ratio $\beta=L/S$, where $L(S)$ is the length of the longest (shortest) side of the top mesa contact, to the potential deformation ratio $\delta=\omega_y/\omega_x$. Breakdown of Hund's first rule was also found in QDs with a deformation ratio $\delta$ as small as 1.2 \cite{4}. Evolution of the excitation spectra in circular and elliptic QDs in a magnetic field up to 4 T was also studied \cite{8}. Whilst the previous works gave some understanding and clarification of the shell structure evolution in response to potential deformations, the detailed analysis and comparison with experimental data is still lacking.

In this work we try to remedy this shortcoming and present calculations of the zero field addition energy spectra and charging diagrams in an applied magnetic field for {\it real} circular and rectangular vertical QD mesa structures. We perform a full scale three-dimensional (3D) analysis of the QDs without any {\it \'a priori} assumptions about the shape or strength of the confinement potential. We utilize density-function theory (DFT) to describe the many body interactions amongst the electrons in the QD, and employ a semi-classical Thomas-Fermi approach to treat charge evolution outside the QD region. In this multi-scale approach, the confinement potential is obtained directly from the self-consistent solution of the Poisson and Schr\"odinger (Kohn-Sham) equations with corresponding device boundary conditions. Our calculations are based on real experimental structures for which well resolved addition energy spectra and magnetic field characteristics have already been measured \cite{3,4,4a}. Previously this full scale method was applied to a variety of problems such as, {\it e.g.}, the simulation of four-gated QD mesas with a square geometry \cite{10} which showed that elliptical deformations have little effect on the addition energy spectra but affected the charge configuration greatly. Preliminary calculations for rectangular QDs also showed that the shell structure is completely destroyed for $\beta>1.2$ and a Wigner molecule in the two-electron system is formed for $\beta>3$ \cite{10a}.

The paper is organized in the following way. The QD structures utilized in this work are briefly described in the next Section. A detailed description of our model is given in Section \ref{Model}. In Section \ref{Results} we first compare addition energy spectra and shell structures of the circular and rectangular QDs in the absence of a magnetic field (Section \ref{B0}). Charging processes and evolution of spin states under application of magnetic field are then considered in Section \ref{Bn0}. Comprarison with experimentally measured charging diagrams in a magnetic field is also performed in the same section. The particularly interesting and important case of a two electron system is addressed in Section \ref{ST}. Finally, Section \ref{Concl} contains concluding remarks.

\section{Device structures}
\label{Structure}

Our QD structures have a layered profile in the vertical $z$-direction \cite{3} which are the same for both the circular and rectangular mesa structures. Two layers of Al$_{0.22}$Ga$_{0.78}$As are used as potential barriers for electrons confined in a quantum well made of In$_{0.05}$Ga$_{0.95}$As. Confinement of electrons in the lateral $xy$-plane is achieved by applying a bias $V_G$ to the gate on the side of the mesa. Full details on the geometry, layer composition of these structure, and the measurement techniques can be found elsewhere~\cite{3,4,4a,13}.

For the simulations described below, we represent the mesa and substrate as a cylinder or a rectangle with total height of 350~nm. The side gate width is taken to be equal to 150~nm, while the center of the quantum well is located 116.5~nm from the top of the side gate. The lower 100~nm of the simulation cylinder is assumed to be buried in the substrate. The diameter of a circular mesa is equal to 0.5 $\mu$m in accordance with experimentally measured mesa \cite{3}. For the rectangular QD mesas, $L$ and $S$ are taken to be 0.55 $\mu$m and 0.4 $\mu$m ($\beta=1.375$), and 0.65 $\mu$m and 0.45 $\mu$m ($\beta=1.44$). These mesas correspond respectively to structures identified as rectangular QD X and rectangular QD Y studied earlier at low magnetic fields \cite{4}. Accordingly we refer to them as rectangular QD X and rectangular QD Y in this paper.

\section{Model and simulation details}
\label{Model}

Within the spin-density functional theory, which we use to describe electrons confined in the QD region, the charge density is calculated after solving separate single-particle Kohn-Sham equations for electrons with spin up ($\uparrow$) and down ($\downarrow$):
\begin{equation}
\hat{H}^{\uparrow(\downarrow)}\psi^{\uparrow(\downarrow)}({\bf r})=\varepsilon^{\uparrow(\downarrow)}\psi^{\uparrow(\downarrow)}({\bf r}),
\label{KSE}
\end{equation}
where the Hamiltonian $\hat{H}^{\uparrow(\downarrow)}$ is given as follows:
\begin{equation}
\hat{H}^{\uparrow(\downarrow)}=\hat{T}+\phi({\bf r})+\Delta E_c({\bf r})+v_{xc}^{\uparrow(\downarrow)}({\bf r}).
\label{H}
\end{equation}

In this equation $\hat{T}$ is the kinetic energy operator which in the presence of an external magnetic field ${\bf B}$ reads as:
\begin{eqnarray}
\hat{T}=\frac{1}{2}\left(-i\hbar\nabla+\frac{e}{c}{\bf A}\right)\frac{1}{m^*({\bf r})}\left(-i\hbar\nabla+\frac{e}{c}{\bf A}\right),
\end{eqnarray} 
where $m^*({\bf r})$ is the position dependent electron effective mass, and ${\bf A}=(1/2)(By,-Bx)$ is the corresponding vector potential. We assume here that the magnetic field is oriented along the $z$-direction and neglect Zeeman splitting for clarity.

In Eq. (\ref{H}), $\Delta E_c({\bf r})$ stands for the conduction band offset between the different materials. The values of $\Delta E_c({\bf r})$ are fixed at 180 and -40 meV for Al$_{0.22}$Ga$_{0.78}$As/GaAs and In$_{0.05}$Ga$_{0.95}$As/GaAs respectively. Note that for the real structures with non-zero doping in the GaAs contact regions, the value of $\Delta E_c({\bf r})$ for In$_{0.05}$Ga$_{0.95}$As/GaAs is difficult to determine precisely.

The exchange-correlation potential $v_{xc}^{\uparrow(\downarrow)}({\bf r})$ is computed within the local spin density approximation (LSDA) \cite{11}. Note that this functional does not explicitly depend on magnetic field. Comparison of DFT results with calculations using current-spin density functional theory \cite{12} for two-dimensional systems showed that this approximation is reliable over a wide range of magnetic field, although at higher fields, effects of paramagnetic currents in $v_{xc}^{\uparrow(\downarrow)}({\bf r})$ become more important.

The potential $\phi({\bf r})=\phi_{ext}({\bf r})+\phi_{ion}({\bf r})+\phi_H({\bf r})$ in Eq. (\ref{H}) is the sum of the external potential $\phi_{ext}({\bf r})$ due to the applied voltage, screening potential $\phi_{ion}({\bf r})$ arising from the ionized impurities in the structure, and Hartree potential $\phi_H({\bf r})$ accounting for the repulsive electron-electron interactions. It is obtained from the solution of the Poisson equation:
\begin{equation}
\nabla\epsilon({\bf r})\nabla\phi({\bf r})=4\pi\rho({\bf r}),
\label{PE}
\end{equation} 
where $\epsilon({\bf r})$ is the dielectric constant of the medium, and $\rho({\bf r})$ is the charge density which inside the QD region is equal to
\begin{equation}
\rho({\bf r})=e\left(\sum_{occup}\left|\psi^{\uparrow}({\bf r})\right|^2+\sum_{occup}\left|\psi^{\downarrow}({\bf r})\right|^2\right),
\end{equation}
with the summations spanning occupied states for electrons with spin up and down (the number of those states is, in general, different). Outside the region, charge distribution is determined from electron $n({\bf r})$ and hole $p({\bf r})$ densities calculated within the semi-classical Thomas-Fermi approximation \cite{13a,13} screened by the ionized donors and acceptors with concentrations  $N_D^+({\bf r})$ and $N_A^-({\bf r})$ (which were assigned their nominal values \cite{3,4,4a}):
\begin{equation}
\rho({\bf r})=e\left[N_D^+({\bf r})-N_A^-({\bf r})+p({\bf r})-n({\bf r})\right].
\end{equation} 

Since vertical QDs are usually much smaller than the physical dimensions of the device, the Kohn-Sham wavefunctions actually vanish long before reaching the device boundaries. This allows us to embed a local region in the global mesh for solving the Kohn-Sham equations. This local region is chosen to be large enough to ensure vanishing wavefunctions on its boundaries.
For the Poisson equation (\ref{PE}), zero normal electric field on the lower part of the simulated structure buried in the substrate and on the top contact region are used as a boundary condition. On other surfaces, not covered by the side gate, the potential $\phi$ is set equal to the Schottky barrier value $V_S = 0.9$ eV. Boundary values of the potential on the side gate are equal to the Schottky barrier value modified by the applied gate bias, $V_S-V_G$. The system of Kohn-Sham and Poisson equations (\ref{KSE}), (\ref{PE}) is solved iteratively until a self-consistent solution for the Kohn-Sham orbitals $\psi^{\uparrow(\downarrow)}({\bf r})$ and eigenvalues $\varepsilon^{\uparrow(\downarrow)}$ is obtained.

The calculations are performed on a parallel platform by means of the finite element method (FEM) with trilinear polynomials on a variable size grid \cite{14}. Previously FEM was applied in electronic structure calculations \cite{14a}. The advantages of FEM utilization are the ability to systematically improve the accuracy by expanding the basis set and its perfect suitability for parallelization. 

In our approach Poisson's equation is solved by means of the damped Newton-Raphson method while the generalized eigenvalue problem obtained after discretization of the Kohn-Sham equation is tackled by means of a subspace iteration method based on a simple Rayleigh-Ritz analysis \cite{14}. The small number of required eigenpairs (in the present work, less than 10) made this approach sufficient. A parallel conjugate-gradient method preconditioned with block Jacobi with an incomplete LU factorization on the blocks available in PETSc (Portable, Extensible Toolkit for Scientific Computation) \cite{15} is utilized for solving the resulting matrix equations. In the presence of a magnetic field, the matrix obtained from the Kohn-Sham equation is Hermitian and the hermitian-conjugate method \cite{15} with the same preconditioner as above had to be used in solving the eigenvalue problem. Compared to the ordinary conjugate gradient method with a Jacobi preconditioner, we found that this approach gives rise to at least an order of magnitude increase in performance especially when dealing with Hermitian matrices \cite{16}. 

After the eigenvalues are determined, the electron charging diagram can be calculated. The experimentally measured charging diagram is obtained by single electron transport spectroscopy \cite{3,4,4a}, when the QD is in the regime of Coulomb blockade with only a very small bias applied across the QD in the vertical $z$-direction. Consequently, the difference between the chemical potentials of the source and drain contacts is negligible and both of them can be set to zero without loss of generality. A charging event occurs when the chemical potential of the QD becomes equal to that of the source (or drain) contact. Then we can determine the stable electronic configuration by directly comparing the total energy $E(N)$ of the $N$-electron system with the system containing $N-1$ electrons, the difference of which gives the required value of the chemical potential $\mu(N)$. Furthermore, we use Slater's rule \cite{17,17a} (transition state technique) in order to calculate $\mu(N)$:
\begin{equation}
\mu(N)=E(N)-E(N-1)\approx\varepsilon(N-1/2),
\end{equation} 
where $\varepsilon(N-1/2)$ is the eigenvalue of the state with half occupancy in the system with $N-1/2$ electrons (the transition state). If $\mu(N)<0$, then the $N$-electron configuration is stable, otherwise the number of electrons in the QD is $N-1$. The voltage $V_G(N)$ at which $\mu(N)=0$ gives the charging voltage for the $N$-th electron (or equivalently it defines the boundary between stable configurations with $N$ and $N-1$ electrons).
 
As mentioned in Section \ref{Intro}, an external magnetic field can induce changes in the spin configuration even when the number of electrons confined in the QD remains constant. Slater's rule can be used only when there is a change in occupancy of {\it one} eigenlevel. If a change in the number of electrons is accompanied by spin rotations (that is, the occupancy of several levels is changed at the same time), it becomes necessary to use several transition states sequentially (this increases the amount of computations but does not invalidate Slater's rule). However, the number of such situations is small so that the overall application of Slater's rule significantly decreases the amount of computational time required to calculate the full charging diagram in magnetic fields 
\footnote{ Our previous calculations \cite{13} have established the validity of Slater's rule for the number of electrons $N>1$. Nevertheless, for completeness we apply Slater's rule to determine the charging point for the first electron as well.}.

Having obtained $\mu(N)$, we can determine the electron addition energy by first computing the so-called capacitive energy \cite{18}:
\begin{equation}
\Delta_2(N)=\mu(N)-\mu(N-1).
\label{eq:delta}
\end{equation}
We evaluate this quantity at the bias $V_G(N)$ corresponding to the addition of the $N$-th electron to the system so that $\Delta_2(N)=-\mu(N-1)$. Together with the charging voltage $V_G(N)$, it is this quantity which is analyzed in detail in the following Section. Note also that this definition of the addition energy is consistent with other calculations where a rigid external potential is used to create electron confinement \cite{3a,4,6,8,10}.

\section{Results and discussion}
\label{Results}

\subsection{Addition energy spectra at zero magnetic field}
\label{B0}

\begin{figure}[p]
\includegraphics[width=8cm]{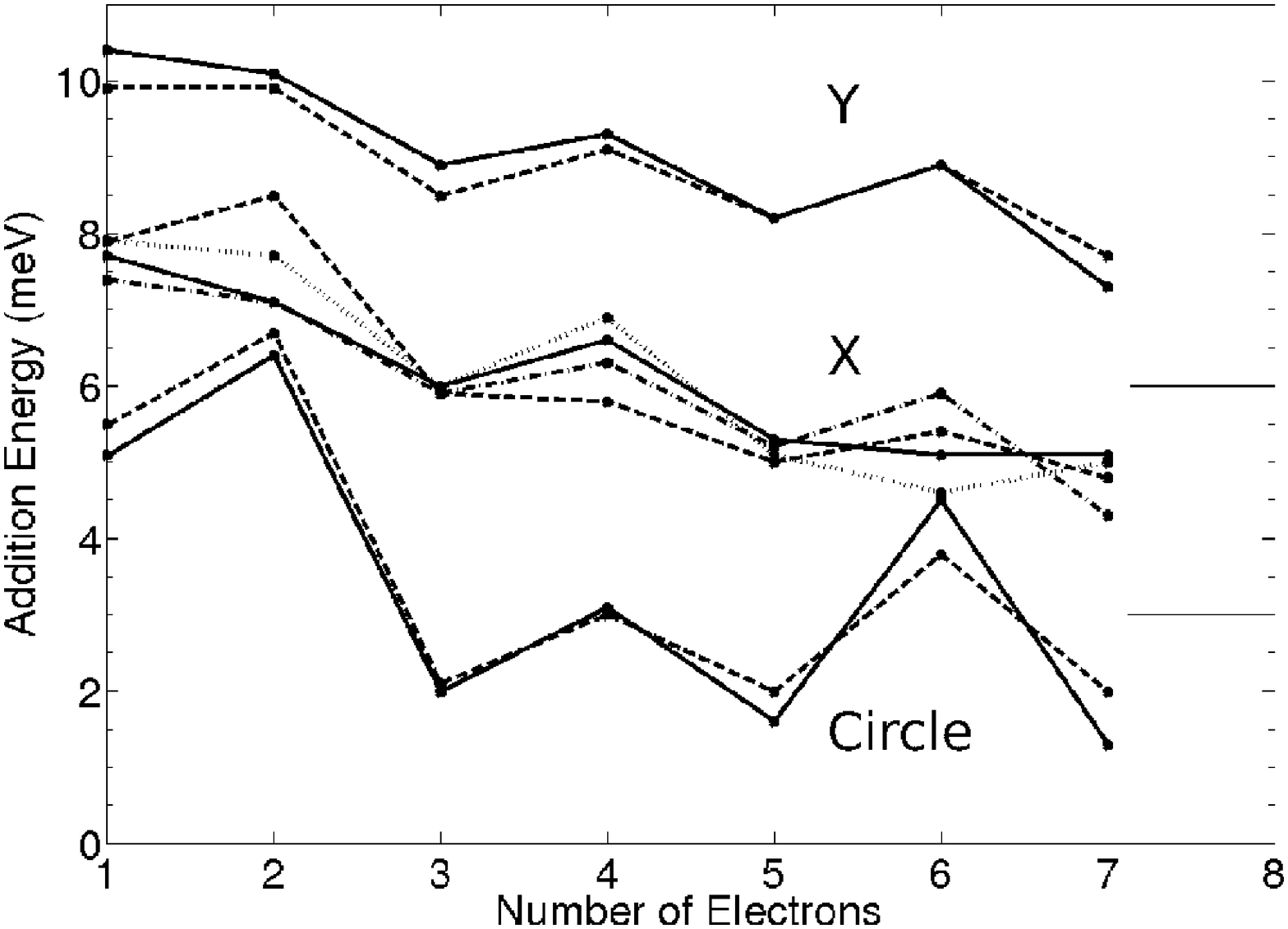}
\caption{Addition energy spectra for the circular QD (lower part), the rectangular QD X (middle part), and the rectangular QD Y (upper part) at zero magnetic field. The parts are shifted by 3 meV in the vertical direction for clarity. In all parts solid lines represent results of calculations, while dashed lines stand for experimental data. Dotted and dash-dotted lines for rectangular QD X are the results of calculations for structures with $\beta=1.306$ and $1.444$ respectively. The calculated curve for the rectangular QD Y is the same as the dash-dotted curve for the rectangular QD X.
\label{fig:add_WXY}
}
\end{figure}

We first consider the addition energy spectra of the circular and rectangular QDs in the absence of a magnetic field. The calculated spectra are shown in Fig. \ref{fig:add_WXY} together with the corresponding experimental spectra taken from Ref. \onlinecite{4}. The spectrum of the QD in the circular mesa (Fig. \ref{fig:add_WXY}, lower part) has been discussed in great detail earlier \cite{4} so here we just briefly review it. It exhibits pronounced maxima for two and six electrons due to the first and second shell closures characteristic of QDs with parabolic 2D circular confinement. The peaks and valleys are a consequence of the interplay between confinement and many-body effects. For two electrons in the QD, the lowest single-particle state $\varepsilon_1$ is fully occupied. The third electron populates the next available eigenstate with $\varepsilon_2>\varepsilon_1$, thus making addition of the electron much more energetically costly than for the case of the second electron. The same situation is repeated for $N=6$ when the second shell is complete. A smaller peak at $N=4$ is due to the fulfillment of Hund's first rule. The total spin $S$ of the $N$-electron system is equal to zero (singlet) for $N=2$ and 6 and to one (triplet) for $N=4$. The agreement between the calculated and experimental spectra is very good, for both peaks and valleys. From the energy separation between the two lowest adjacent eigenvalues we can deduce that the confinement strength $\hbar\omega$ in the empty QD is about 6 meV.

The analysis for the rectangular QD spectra is more challenging. From a simple model of a 2D anisotropic harmonic oscillator with eigenenergies $\varepsilon_{n_x,n_y}=\hbar\omega_x(n_x+1/2)+\hbar\omega_y(n_y+1/2)$, with quantum numbers $n_x,~n_y=0,~1,~...$, one can see that in general, except at certain special deformations, there are no degeneracies in the eigenspectrum at 0 T \cite{4}. This means that the shell structure of a circular QD with strong maxima for 2, 6, 12, ... electrons in the addition energy spectra will not hold here, and one should expect maxima for most even electron-numbers. This simple interpretation is largely confirmed by our calculations. 

Experimentally, because of side wall etching, the sizes of the fabricated mesas can be determined only with about 5~\% accuracy \cite{4}. To assess the effect of this uncertainty for the rectangular QDs, we first performed calculations of the addition spectra for three systems with geometrical aspect ratios equal and $\pm 5~\%$ of the value given for rectangular QD X: namely $\beta=1.306$, $1.375$ and $1.444$. Note that the last structure has an aspect ratio that is very close to the nominal value $\beta=1.44$ for rectangular QD Y \cite{4}. Hence, we compare our calculated spectra to spectra for both rectangular QDs X and Y. The results are also given in Fig. \ref{fig:add_WXY}, middle and upper part respectively. Surprisingly our calculations for $\beta=1.444$ are in excellent agreement with the experimental data for rectangular QD Y, whilst none of the 5~\% variation-spectra are able to describe the experimental results as well for rectangular QD X, although the calculated spectrum for the structure with $\beta=1.444$ is closest to that measured. In fact, the experimental addition spectrum for X is similar in some respects to the one for the circular QD indicating that this structure may have an effective aspect ratio much lower than can be assumed on the basis of its geometrical aspect ratio alone. This conclusion is also corroborated by examining the magnetic field dependence of the charging digrams for these structures (see Section \ref{Bn0}).

Further inspection of the calculated spectra reveals that the difference (contrast) between the peaks and valleys for the rectangular QDs is much lower than for the circular QD. We also note that the peak for $N=2$ is significantly suppressed, and a 10~\% difference in $\beta$ gives rise to the $N=6$ feature changing from a minima to a maxima (increasing $\beta$ from 1.306 to 1.444). This shows that the electronic structure in rectangular QDs is very sensitive to even small distortions of the confinement potential and the precise balance between all contributions to the total energy is essential for its accurate description. On the other hand, the four electron peak appears to be robust and is present in all the calculated addition energy spectra. This is because there is a trivial shell closure at $N=4$ since there is now no degeneracy between the second and third lowest single-particle eigenvalues as is the case for the circular QD. 

\begin{figure}[p]
\includegraphics[width=8cm]{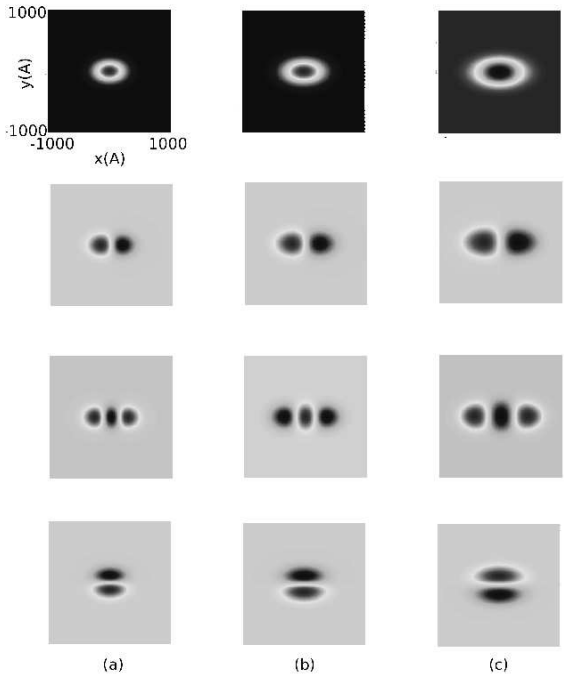}
\caption{Two-dimensional contour plots of the wave-functions for the four lowest eigenvalues (from top to bottom) in the rectangular QD with $\beta=1.444$: (a) $N=0$, $V_G=-1.6$ V, (b) $N=0$, $V_G=-1.3$ V, (c) $N=6$, $V_G=-1.3$ V. Coordinate $z$ is fixed in the center of the QD. \label{fig:evect_all}
}
\end{figure}

In order to better understand the electronic properties of a rectangular QD, we plot the wave-functions for the four lowest eigenvalues for the case of an empty QD at $V_G=-1.6$ and $-1.3$ V, and also for the QD containing six electrons at $V_G=-1.3$ V (Fig. \ref{fig:evect_all}). One can see that the "spread" of the wave-functions progressively increases from (a) to (c) not only because the presence of electrons in the QD induces a repulsive Coulomb potential which effectively "flattens" the bottom of the conduction band, but also this effect occurs in the empty QD by solely varying the gate bias. This is a consequence of the three-dimensionality of the mesa structure, which cannot be reproduced in model 2D calculations with a rigid confinement potential. The doped regions just above and below the QD are never fully depleted, and changes in the gate bias affect the charge distribution in them as well. The resulting changes in the potential are carried over into the QD and lead to the effective decrease of the confinement strength as the gate bias becomes more positive. The same effect is also present in circular QDs \cite{10}. It can be seen that the confinement decreases with increasing gate bias and the population in the QD. At the same time, the potential deformation ratio $\delta$ also increases, {\it i.e.} the QD becomes more elongated. This is a manifestation of a self-consistent effect due to which electrons tend to localize as far as possible from each other in order to minimize the Coulomb repulsion. This results in a more effective decrease of the confinement along the $x$-direction than the $y$-direction.

\subsection{Charging diagrams in a magnetic field}
\label{Bn0}

\begin{figure}[p]
\includegraphics[width=8cm]{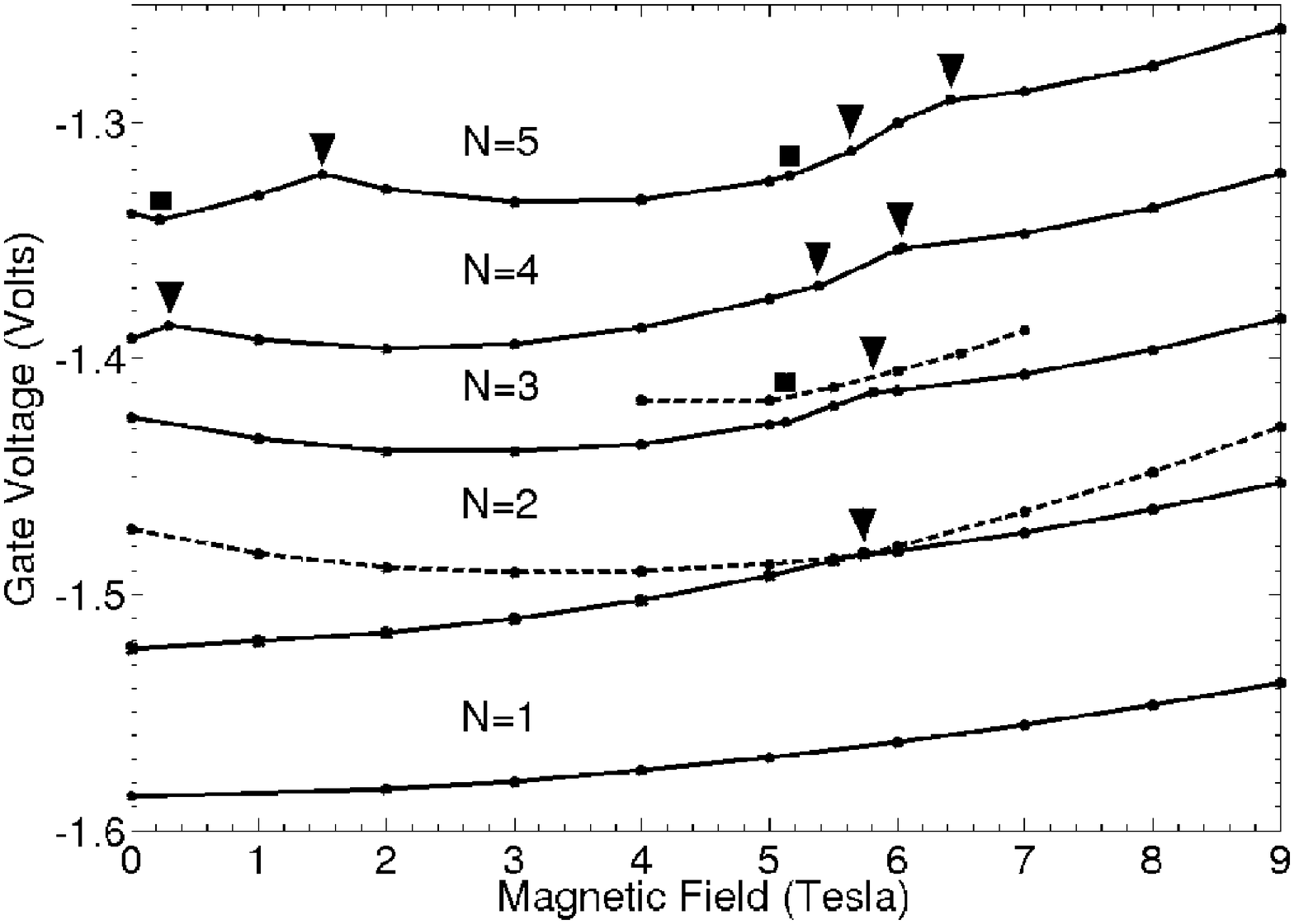} 
\caption{Calculated charging diagram (solid curves) for ground states in a magnetic field for the circular QD. The dashed lines show some excited states. Triangles ($\blacktriangledown$) define boundaries of regions with different total spin and angular momentum in the ground state of the $N$-electron system. Squares ($\blacksquare$) show features in the ground state of the $N$-electron system due to reconstruction of the $(N-1)$-electron system.
\label{fig:newtar1}
}
\end{figure}

In Fig. \ref{fig:newtar1} we plot the calculated charging voltage for $N=1$ to $N=5$ electrons as a function of magnetic field (charging diagram) for the circular QD. Again these results have already been extensively discussed elsewhere \cite{3a,19} but they are used here to compare the properties of circular and rectangular QDs. Comparison with experiment for the circular QD \cite{4a} shows that overall agreement is very good, albeit the confinement potential at zero field in our model structure is somewhat weaker, {\it i.e.}, the curves are more closely spaced. The charging voltages generally (in the considered range) increase with magnetic field since the effective confinement becomes stronger. The curves corresponding to the charging of $N\ge 2$ electrons exhibit "cusps" corresponding to various magnetic field induced spin and angular momentum transitions. For the $N=3$ charging curve arising from the addition of the third electron to the two-electron system, the cusp near 5 T (depicted by a square) is due to a change in the ground state configuration of the two-electron system at $B\approx 5.6$ T (the singlet-triplet transition) which affects the addition energy of the third electron. The shift from 5.6 T to 5.0 T is due to a screening effect and is consistent with the experiment \cite{4a}. The cusp near 6 T (triangle) reflects an increase in the total spin of the three-electron system from $S=1/2$ to $S=3/2$, namely below this point two electrons are spin-up and one is spin-down while above this point all three electrons in the QD are spin-up and form a spin-polarized system. The dashed line in the 4 to 7 T region shows an excited state where three electrons have a total angular momentum $L=2$ but total spin $S=1/2$. In our calculations this state never becomes the ground state, which is different from recent results based on the configuration interaction method \cite{19} where it was found that the transition from the ground state ($L=1$, $S=1/2$) to this state ($L=2$, $S=1/2$) was responsible for the lower field cusp. However, the predicted difference in the charging voltages is so small that given the large number of experimental parameters (such as dopant distribution, geometrical size of the structure, conduction band offsets, etc.) whose values cannot possibly be determined precisely, the real cause for this feature is difficult to ascertain at best.

In general, the rightmost cusp in the Fig. \ref{fig:newtar1} charging diagram always corresponds to complete spin polarization of the electron system. The magnetic field at which the formation of this state occurs increases with the number of electrons since a stronger field is required to overcome the large kinetic energy accompanying single occupancy of consecutive orbitals. Cusps in the $N$-electron curves in the vicinity of $B=5$ T are either due to changes in the $(N-1)$-electron configuration or due to single spin rotations in the $N$-electron system. The two cusps around $B\approx 0.25$ T in the $N=4$ and $N=5$ curves mark the breakdown of Hund's first rule filling in the four-electron system and respective change in the addition energy of the fifth electron due to the decrease of exchange energy in the four electron system, similar to the $N=3$ curve. The cusp at $B\approx 1.5$ T in the $N=5$ curve can be understood in terms of the Fock-Darwin spectrum describing non-interacting particles \cite{3a}. Around that point, an electron undergoes a transition in the quantum numbers moving from the Fock-Darwin state with $(n,l)=(0,-1)$ to $(n,l)=(0,2)$, thereby increasing the total angular momentum of the system while keeping the total spin value constant.

\begin{figure}[p]
\includegraphics[width=8cm]{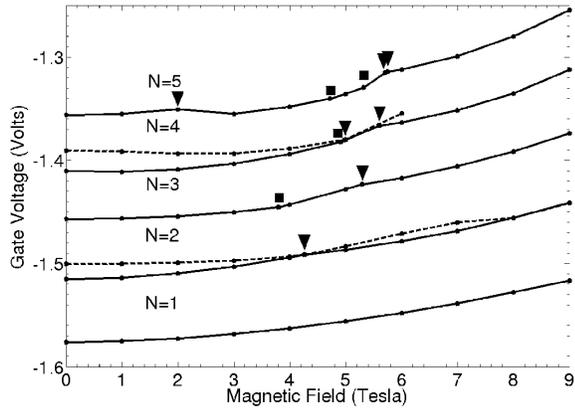}
\caption{Same as in Fig. \ref{fig:newtar1} but for the rectangular QD with $\beta=1.444$. In this case triangles ($\blacktriangledown$) define boundaries of regions with different total spin and ($n_x$, $n_y$) quantum numbers in the $N$-electron system.  
\label{fig:add_X2}
}
\end{figure}

The calculated charging diagram of the rectangular QD with $\beta=1.444$ shown in Fig. \ref{fig:add_X2} superficially looks similar to that of the circular QD. However, all cusps are less pronounced and the separation between consecutive curves is generally smaller. The former is due to the absence of a clear shell structure while the latter finds its origin in the decreased separation between the lower eigenlevels arising from the weaker confinement in the $x$-direction.
We also observe that the magnetic fields at which transitions occur are approximately the same as in the circular QD (compare with Fig. \ref{fig:newtar1}). This can be explained by noticing that the QD surface area and the resulting electron density in both QDs are approximately equal. Since in the DFT the electron density determines the magnitude of the Coulomb and exchange interactions, it takes roughly the same amount of energy in the two QD systems to change these interactions in order to induce single electron transitions. The charging curves move upward slightly faster with magnetic field than those in Fig. \ref{fig:newtar1}. The reason for this behavior is again, compared to the magnetic confinement $\omega_c=eB/mc$, a smaller value of $\hbar\omega_x$ so that the magnetic field affects the related energy levels more strongly. Note that for the rectangular QD the $N=4$ curve corresponds, at zero field, to the charging of the fourth electron to form a singlet $S=0$ state (two electrons with spin-up, two with spin-down), and this behavior is clearly different from that for the circular QD where four electrons obey Hund's first rule close to zero magnetic field. The singlet state continues to be the ground state up to $B\approx 5$ T where the $S=1$ state (shown by a dashed curve) briefly takes over. 

\begin{figure}[p]
\includegraphics[width=8cm]{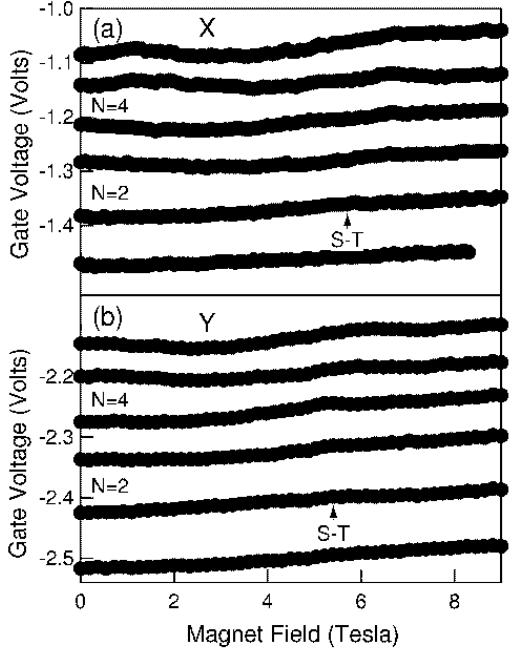}
\caption{Experimental charging diagrams showing ground state evolution with magnetic field for (a) rectangular QD X, and (b) rectangular QD Y. The position of the $N=2$ singlet-triplet (S-T) transition is marked. 
\label{fig:Fig280105-1}
}
\end{figure}

At low magnetic fields (but above $B\approx 0.25$ T), the calculated charging voltages for both $N=3$ and $N=4$ in the circular QD (Fig. \ref{fig:newtar1}) clearly decrease whilst the corresponding curves for the rectangular QD are initially "flat" before moving upward (Fig. \ref{fig:add_X2}). This is due to the different response of the second lowest single-particle state to the magnetic field. For a circular potential it first decreases and then increases with magnetic field, whilst a sufficiently large potential deformation results in the monotonic increase of this eigenvalue with magnetic field. This is what happens in our modelled structure where $\delta\gtrsim 2$ (see Fig. \ref{fig:evect_all}). At small magnetic fields the experimental charging diagram for rectangular QD X [see Fig. \ref{fig:Fig280105-1}(a)] actually looks very similar to both the experimental and calculated diagrams for a circular QD (see Ref. \onlinecite{4a} and Fig. \ref{fig:newtar1} respectively), although notably the cusps related to Hund's first rule filling for $N=$ 4 are absent near 0 T. This indicates that the effective aspect ratio $\delta$ characterizing the confinement potential in rectangular QD X is much closer to one than could be deduced from its geometrical aspect ratio $\beta$. On the other hand, the measured charging diagram for rectangular QD Y [Fig. \ref{fig:Fig280105-1}(b)] is closer in appearance to the calculated diagram for the rectangular QD shown in Fig. \ref{fig:add_X2}. In particular, curves for $N=3$ and $N=4$ move gradually upwards, whilst the curve for $N=5$ moves only slightly downwards. The experimental charging spectra for both circular and rectangular QDs nonetheless reveal a weaker dependence of the charging voltages on the magnetic field and a larger separation between consecutive curves (within 30\%) not observed in the theoretical spectra 
\footnote{ Surprisingly the smaller separation between the calculated curves does not affect the addition energy spectra which are in good agreement with measured data (Section \ref{B0}). This is because the addition energy is defined not only by the gate voltage difference between consequitive charging points but rather by the slope of $\mu(N,V_G)$ vs. $V_G$ which in itself is a highly non-trivial function of the QD device parameters and electron number (for discussion, see, {\it e.g.,} [\onlinecite{13,19}] and references therein).}. 
Both these discrepancies may be explained if the confinement is indeed stronger (in both $x$ and $y$ directions) in the experimental devices, so that the magnetic field is unable to perturb it appreciably.

\subsection{Singlet-triplet transition in a two electron system}
\label{ST}

\begin{figure}[p]
\includegraphics[width=8cm]{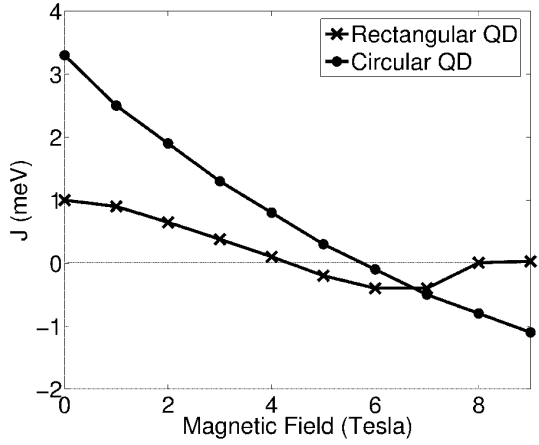}
\caption{Calculated singlet-triplet energy separation $J$ in the two-electron system as a function of the magnetic field. $\beta=1.444$ for the rectangular QD. 
\label{fig:st_X2_W}
}
\end{figure}

One of the most striking differences between the calculated charging behavior of the QDs in the circular and rectangular mesa structures (Figs. \ref{fig:newtar1} and \ref{fig:add_X2}) is found in the two electron system. The dashed curves for $N=2$ in both plots represent excited states. At low magnetic fields, below the cusp on the charging curve, the excited state is a triplet ($S=1$) whilst at high magnetic fields, above the cusp, it is a singlet ($S=0$). This cusp in fact marks the singlet-triplet (S-T) transition. From the full and dashed curves in Figs. \ref{fig:newtar1} and \ref{fig:add_X2}, we can extract (see Eqs. 7 and \ref{eq:delta}) the energy separation between the singlet and triplet states, namely 
\begin{equation}
J=E^T(2)-E^S(2)=\Delta_2^T(2)-\Delta^S_2(2),
\end{equation}
where symbols T and S stand for triplet and singlet states respectively. This singlet-triplet energy difference for a QD with two electrons is very interesting because of its potential importance for quantum computation. To the best of our knowledge this is the first calculation of this important characteristic in a {\it realistic} system taking into account the full 3D nature of the problem. Hitherto, the singlet-triplet separation has been considered in a realistic non-circular QD using 2D confinement potentials produced by the gate voltages only \cite{20a}. Our theoretical data for both circular and rectangular QDs as a function of magnetic field are shown in Fig. \ref{fig:st_X2_W}. One can see that in the range of investigated magnetic fields the behavior of $J$ in the two structures is radically different. In the limit of large magnetic field $J\rightarrow 0$ for the rectangular mesa QD \cite{6a}, whilst it continues to become more negative without any saturation in the case of the circular QD. Also, the value of $J$ for the rectangular QD is about three times smaller than for the circular QD at zero magnetic field.

\begin{figure}[p]
\includegraphics[width=8cm]{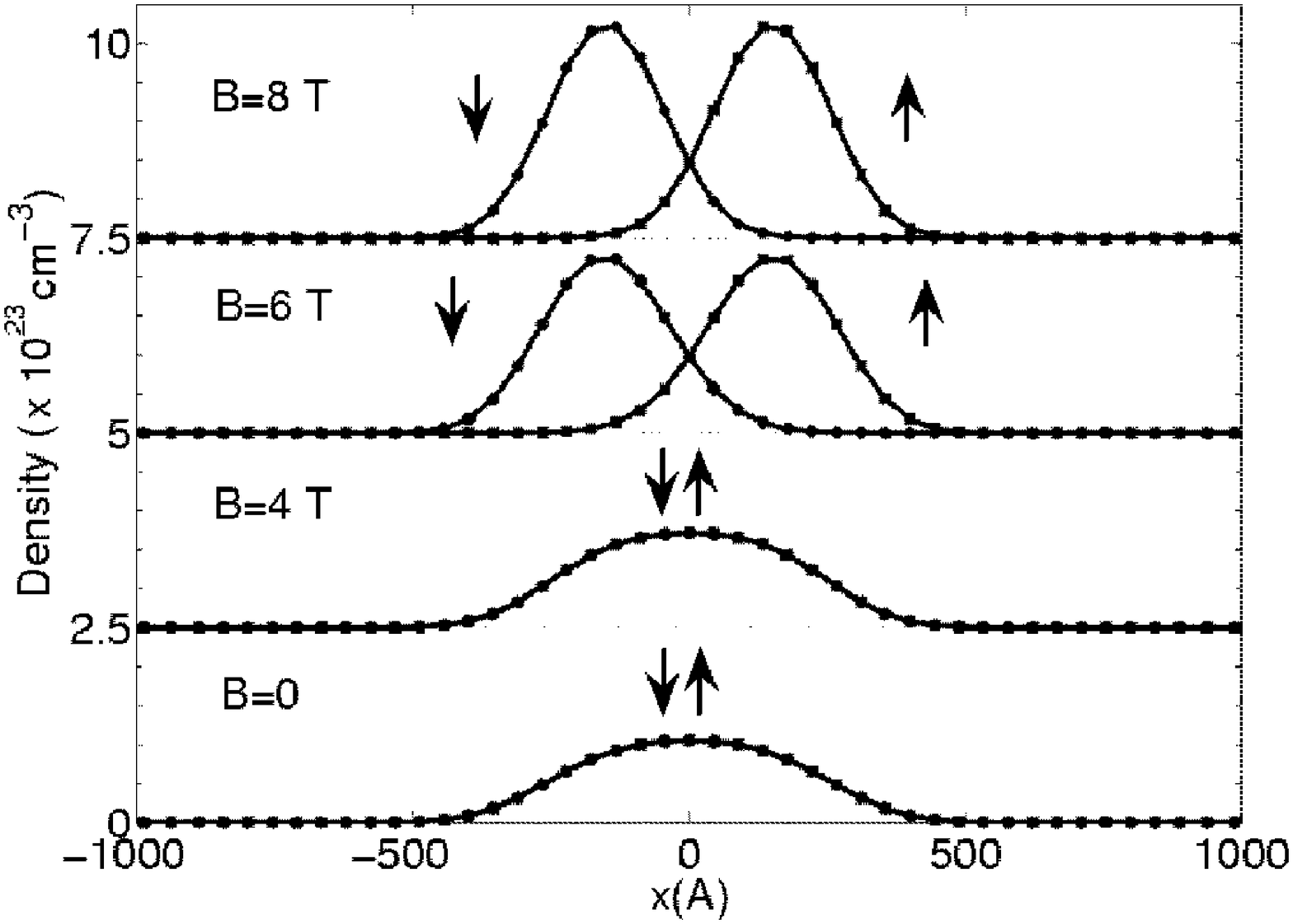}
\caption{Single-particle electron densities in the $x$-direction in the rectangular QD with $\beta=1.444$ for the singlet state at different magnetic fields (coordinates $y$ and $z$ are fixed in the middle of the QD). Arrows ($\uparrow$) and ($\downarrow$) correspond to the electrons with spin up and down. The data for $B=0$, $4$ T and $6$, $8$ T are obtained at $V_G=-1.45$ and $-1.47$ V respectively. The curves for different magnetic fields are shifted in the vertical direction by $2.5\times10^{23}$ cm$^{-3}$ for clarity.
\label{fig:n2_s_all}
}
\end{figure}

\begin{figure}[p]
\includegraphics[width=8cm]{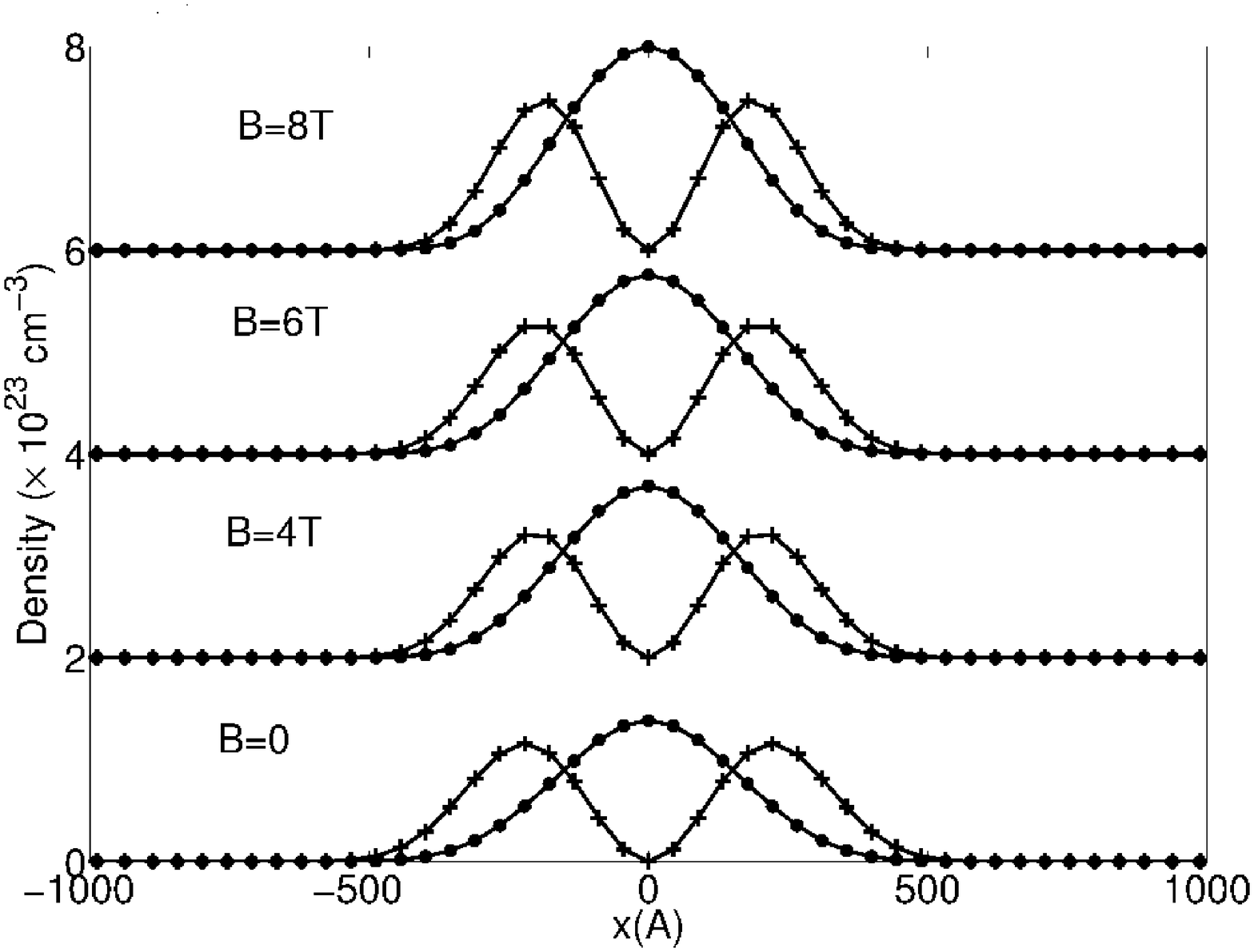}
\caption{Same as in Fig. \ref{fig:n2_s_all} but for the triplet state. Curves with circles ($\bullet$) and pluses ($+$) are for the ground and first excited occupied states respectively. The curves for different magnetic fields are shifted in the vertical direction by $2.0\times10^{23}$ cm$^{-3}$ for clarity.
\label{fig:n2_t_all}
}
\end{figure}

In order to gain a better insight into the calculated behavior of $J$ in a magnetic field, we show in Figs. \ref{fig:n2_s_all} and \ref{fig:n2_t_all} respectively the eigenfunctions of the occupied singlet and triplet states in a rectangular QD with $\beta=1.444$. In Fig. \ref{fig:n2_s_all} we see that at low fields both of the electrons in the singlet state occupy the same space in the QD, {\it i.e.}, their wave-functions are identical, and they become ever more localized at the center of the QD with increasing field. However, above $B\approx 5.8$ T the electrons start to localize in opposite regions of the QD along the $x$-direction inducing a symmetry "breaking" so that the wave-functions no longer have the spatial symmetry of the confinement potential around its minimum at the center of the QD. This effect appears similar to so-called Wigner crystallization, {\it i.e.,} spatial separation of electrons, caused by the applied magnetic field \cite{3a}. Further increase of the magnetic field gives rise to further localization of the electrons. In the case of the triplet state (Fig. \ref{fig:n2_t_all}), there is no such radical reconstruction of the electron wave-functions. They just become more localized around the center of the QD. The symmetry breaking in the singlet state also results in a change of the total electron density. It attains two "humps" similar to the electron density in the triplet state. The reconstruction of the singlet electron density is responsible for the predicted evolution of the singlet-triplet splitting at large magnetic fields. When the electron densities of both states become similar, exchange and Coulomb energies also have close values. Coupled with the fact that the single-particle energy separation decreases with increasing field (the energy levels condense in to the lowest Landau band \cite{3a}), this results in a vanishing $J$.
Interestingly, as can be seen from Figs. \ref{fig:n2_s_all}, the overlap between the electron wave-functions at $B=8$ T is still appreciable (and close in value to that one at $B=6$ T), and yet $J\rightarrow 0$.

The symmetry breaking does not occur in a circular QD where the electron density of the singlet state retains the symmetry of the confinement potential in the range of magnetic fields considered in this work. As a result, the exchange energy of the triplet state is always smaller than that of the singlet. Starting from zero field, the decrease in the separation between the two lowest single-particle energies as the magnetic field is increased leads to the triplet becoming the ground state at $B\approx 5.6$ T, but the singlet-triplet energy difference $J$ does not exhibit any saturation behavior at higher field. 

The behavior of the calculated singlet-triplet separation for the circular QD is in good agreement with experimental data obtained by excited state spectroscopy \cite{4a} which enables the observation of electron tunneling through both ground and excited states. Although there is extensive experimental data for the circular QD, detailed excited state spectra for the vertical rectangular QDs are currently unavailable. Clearly in light of the predictions concerning the behavior of $J$, it would be interesting to perform such experiments in the future. At present, we can only identify the two-electron S-T transtion for both rectangular QDs in the charging diagrams of Fig. \ref{fig:Fig280105-1} which just show the evolution of the ground states. We do, however, note that recently, co-tunneling was utilized to extract the values of the singlet-triplet energy sepration for two electrons in a lateral QD with a spatially elongated (approximately harmonic) confinement potential \cite{20} where it was found that with increasing magnetic field, $J$ changes sign and then saturates, consistent with our calculations.

\section{Conclusion}
\label{Concl}

In this work we studied the electronic properties of vertical quantum QDs in circular and rectangular mesas using a multiscale density-functional approach within the local spin density approximation. The QD was treated fully quantum mechanically and the surrounding regions were described by a semiclassical Thomas-Fermi approximation. The calculations were done on a 3D mesh by means of a finite element method. Slater's rule (the transition state technique) was utilized to study charging of the QDs. With this approach, we were able to calculate the zero magnetic field electron addition energy spectra in the QDs as well as study the electron charging diagrams in a magnetic field. 

We found that the zero field addition energy spectrum and charging diagram (charging voltages as a function of magnetic field) of a circular QD are in good agreement with available experimental results. For rectangular QDs, we found that a 5~\% variation from the nominal aspect ratio can give rise to marked differences in the addition energy spectra. On the basis of a comparison with experimental data in a magnetic field for rectangular mesa structures (Rectangles X and Y in Ref. \onlinecite{4}) with different aspect ratios, we conclude that rectangular QD X most likely has a smaller effective aspect ratio than previously reported, whilst our calculated results are in good agreement with those obtained for the rectangular QD Y.

From the charging diagrams, we also extracted the calculated magnetic field dependence of the singlet-triplet energy separation $J$ for a two electron system. Up to the singlet-triplet transition, this quantity decreases with magnetic field for both circular and rectangular QDs. However, beyond that transition, in the limit of a strong field, $J$ for the rectangular QD asymptotically approaches zero whilst it continues to become more negative in the case of the circular QD in the range of considered magnetic fields. Analysis of electron wave-functions revealed that the behavior of the singlet-triplet energy separation in a rectangular QD can be attributed to the "breaking" of the symmetry of the singlet wave-functions.

\acknowledgments

This work has been supported by DARPA QUIST program DAAD19-01-1-0659 and Strategic Application Program (SAP) at NCSA.

\begin{thebibliography}{100}

\bibitem{1} K.K. Likharev, Proc. IEEE {\bf 87}, 606 (1999).

\bibitem{2} D. Loss and D.P. DiVincenzo, Phys. Rev. A {\bf 57}, 120 (1998).

\bibitem{3} S. Tarucha, D.G. Austing, T. Honda, R.J. van der Hage, and L.P Kouwenhoven, Phys. Rev. Lett. {\bf 77}, 3613 (1996).

\bibitem{3a} S.M. Reimann and M Manninen, Rev. Mod. Phys. {\bf 74}, 1283 (2002).

\bibitem{3b} P. Matagne, J.-P. Leburton, D.G. Austing, and S. Tarucha, Phys. Rev. B {\bf 65}, 085325 (2002).

\bibitem{4} D.G. Austing, S. Sasaki, S. Tarucha, S.M. Reimann, M.Koskinen, and M. Manninen, Phys. Rev. B {\bf 60}, 11514 (1999).

\bibitem{4a} L.P. Kouwenhoven, T.H. Oosterkamp, M.W.S. Danoesastro, M. Eto, D.G. Austing, T. Honda, and S. Tarucha, Science {\bf 278}, 1788 (1997).

\bibitem{6} A.V. Madhav and T. Chakraborty, Phys. Rev. B {\bf 49}, 8163 (1994); S. Tarucha, T. Honda, D.G. Austing, Y. Tokura, K. Muraki, T.H. Oosterkamp, J.W. Janssen, and L.P. Kouwenhoven, Physica E {\bf 3}, 112 (1998); T. Ezaki, Y. Sugimoto, N. Mori, and C. Hamaguchi, Semicond. Sci. Technol. {\bf 13}, A1 (1998); K. Hirose, N.S. Wingreen, Phys. Rev. B {\bf 59}, 4604 (1999); I.-H. Lee, Y.-H. Kim, and K.-H Ahn, J. Phys.: Condens. Matter {\bf 13}, 1987 (2001); B. Szafran, F.M. Peeters, S. Bednarek, and J. Adamowski, Phys. Rev. B {\bf 69}, 125344 (2004).

\bibitem{6a} P.S. Drouvelis, P. Schmelcher, and F.K. Diakonos, Phys. Rev. B{\bf 69}, 155312 (2004).

\bibitem{8} Y. Tokura, S. Sasaki, D.G. Austing, and S. Tarucha, Physica B {\bf 298}, 260 (2001).

\bibitem{10} P. Matagne, and J.-P. Leburton, Phys. Rev, B {\bf 65}, 155311 (2002).

\bibitem{10a} P. Matagne and J.-P. Leburton, in {\it Quantum Dots and Nanowires}, Eds. S. Bandyopadhyay and H. S. Nalwa, ASP, California (2003).

\bibitem{11} J.P. Perdew and Y. Wang, Phys. Rev. B {\bf 45}, 13244 (1992). 

\bibitem{12} F. Ancilotto, D.G. Austing, M. Barranco, R. Mayol, K. Muraki, M. Pi, S. Sasaki, and S. Tarucha, Phys. Rev. B {\bf 67}, 205311 (2003).

\bibitem{13a} S.M. Sze, {\it Physics of Semiconductor Devices, 2nd Edition}, Wiley Interscience, 1981.

\bibitem{13} P. Matagne, and J.-P. Leburton, Phys. Rev, B {\bf 65}, 235323 (2002).

\bibitem{14} K.-J. Bathe, {\it Finite Element Procedures in Engineering Analysis}, Prentice-Hall, Inc., New Jersey, 1982.

\bibitem{14a} S.R. White, J.W. Wilkins, and M.P. Teter, Phys. Rev. B {\bf 39}, 5819 (1989); J.E. Pask, B.M. Klein, C.-Y. Fong, and P.A. Sterne, {\it ibid.} {\bf 59}, 12352 (1999).

\bibitem{15} S. Balay, K. Buschelman, V. Eijkhout, W.D. Gropp, D. Kaushik, M.G. Knepley, L.C. McInnes, B.F. Smith and H. Zhang, {\it PETSc Users Manual}, ANL-95/11 - Revision 2.1.5, Argonne National Laboratory, 2004. 

\bibitem{16} J. Fettig, D.V. Melnikov, N. Sobh, and J.-P. Leburton (unpublished).

\bibitem{17} J.C. Slater, Adv. Quantum Chem. {\bf 6}, 1 (1972).

\bibitem{17a} L.R.C. Fonseca, J.L. Jimenez, J.-P. Leburton, and R.M. Martin, Phys. Rev. B{\bf 57}, 4017 (1998).

\bibitem{18} I.-H. Lee, K.-H Ahn, Y.-H. Kim, R.M. Martin, and J.-P. Leburton, Phys. Rev. B {\bf 60}, 13720 (1999).

\bibitem{19} B. Szafran, S. Bednarek, and J. Adamowski, Phys. Rev. B {\bf 67}, 115323 (2003).

\bibitem{20a} J. Kyriakidis, M. Pioro-Ladriere, M. Ciorga, A.S. Sachrajda, and P. Hawrylak, Phys. Rev. B {\bf 66}, 035320 (2002).

\bibitem{20} D.M. Zumb\"uhl, C.M. Marcus, M.P. Hanson, and A.C. Gossard, Phys. Rev. Lett. {\bf 93}, 256801 (2004).

\end{thebibliography}
\end{document}